\documentclass[conference, onecolumn]{IEEEtran}
\IEEEoverridecommandlockouts

\usepackage{amsmath,amssymb,amsfonts}
\usepackage{algorithmic}
\usepackage{graphicx}
\usepackage{textcomp}
\usepackage{xcolor}
\usepackage{subcaption}
\usepackage{optidef}
\usepackage[backend=bibtex, sorting=none]{biblatex}
\bibliography{biblio.bib}
\def\BibTeX{{\rm B\kern-.05em{\sc i\kern-.025em b}\kern-.08em
    T\kern-.1667em\lower.7ex\hbox{E}\kern-.125emX}}

\begin{document}

\title{Near-Field Beampointing with Low Exposure Regions: a Dominant Subspace Projection Approach \\
}

\author{\IEEEauthorblockN{
Laurence Defraigne\thanks{\footnotesize{L. Defraigne thanks the FRS-FNRS and the FRIA for their financial support.}}, Gilles Monnoyer, Jérôme Louveaux, Luc Vandendorpe}
\IEEEauthorblockA{\textit{ICTEAM, UCLouvain, Louvain-la-Neuve (Belgium), emails~: firstname.lastname@uclouvain.be} }
}

\maketitle

\begin{abstract}
The spherical nature of the wavefronts exhibited in the near-field of antenna arrays enables advanced beamforming capabilities, such as beampointing and beamnulling. In this paper, we exploit these properties to design a near-field beam pattern under a low exposure region constraint. We address the continuous region constraint through spatial discretization, which results in a large number of constraints that lead to prohibitive computational complexity. We propose a novel low-complexity algorithm that enables a computationally tractable beam pattern design. It uses a low-dimensional subspace representation of the low exposure region based on a singular value decomposition. Our approach achieves low complexity while providing a power received at a target user close to the optimal achievable power, yet with uniform power mitigation over the low exposure region.

\end{abstract}


\section{Introduction}

Future communication networks are expected to feature larger antenna arrays evolving to extra-large MIMO (XL-MIMO) configurations, and higher carrier frequencies extending into the terahertz (THz) range. Both factors increase the so-called Fraunhofer distance, thereby expanding the near-field (NF) region of antenna arrays. In this region, the plane-wave approximation is no longer accurate and the true spherical nature of the electromagnetic waves must be taken into account \cite{NF_param}. Due to the spherical wavefront, beam patterns can be specified in both angle and range. This enables, for example, beampointing, beamnulling and user separation along both spatial dimensions \cite{NF_MU, near_field_nulling}.

This capability can be exploited for a new design that creates low exposure regions (LERs) where the received power is mitigated within a two- or three-dimensional continuous region (CR). Designing beampointers achieving such LERs is essential for protecting sensitive areas, such as hospital rooms, and for preventing communication in specific regions. In addition, at higher frequencies, the beampointing vector must be updated more frequently since the user mobility becomes more critical as the frequencies increase. This thus requires a low-complexity algorithm compatible with real-time computation. 

The beampointing ability in NF has been studied through maximum ratio transmission (MRT), minimum mean square error (MMSE), zero-forcing (ZF) and relaxed zero-forcing (RZF) for interference mitigation in multi-user scenarios \cite{NF_MU}. Since interference amounts to unwanted received power at non-intended locations, it can be formulated as a power suppression problem through constraints on the received power at specified positions. As a result, interference mitigation techniques are directly applicable to LER constraints. More complex beam pattern designs have been explored using techniques such as linear constraint minimum variance (LCMV), particle swarm, and deep neural network techniques \cite{near_field_nulling}. However, these approaches focus on a limited number of discrete and distributed user locations and do not explicitly address interference or power suppression over CRs. Moreover, they present a complexity incompatible with real-time computation.

Low-complexity beampointer designs have been studied such as in \cite{ConvexOpti} where the authors adapt convolutional beamspace techniques to a NF scenario. However, their approach relies on convex solvers that are efficient for a small number of discrete user locations, but become computationally intractable for a higher number of constraints, which could result from the discretization of a CR. Authors in \cite{RZF_SOPC} propose a low-complexity RZF algorithm designed for scenarios with a large number of interference constraints, which could be applied to a sampled LER. However, their algorithm assumes linearly independent interference channels and does not exploit the NF structure of the channels, making it unsuitable for a large number of constraints representing a sampled compact region. Despite these differences, we include \cite{RZF_SOPC} as a baseline for comparison with our method.

Interference suppression over a CR has been studied in the case of location uncertainties, such as in \cite{Interfer_Region_average_FF} where the power constraint is averaged over the uncertainty region, but only for an angular CR in FF, which is thus not applicable to an LER constraint in NF beampointing. The challenge of a CR constrained beam pattern design also arises in physical layer security to address the uncertainty on an eavesdropper position, such as in \cite{Secure_Uncertain_FF} but only in FF. NF scenarios are also studied in \cite{Secure_max_SNLR}, but the secrecy metrics are optimized for eavesdroppers located anywhere outside the receiver uncertainty region, instead of suppressing power within a specified compact CR. Consequently, these approaches cannot be directly applied to an LER constrained beam pattern design.

To the best of the authors’ knowledge, existing NF beam pattern designs do not provide power suppression over a CR with low complexity suitable for tractable operation. This work addresses this gap by proposing a low-complexity NF beam pattern design that creates a two-dimensional LER while maximizing the user received power.

To summarize, the contributions of this paper are as follows:
\begin{itemize}
    \item We propose a low-dimensional representation of the LER, capturing all the structure of the CR.
    \item We propose a new algorithm that, unlike existing methods, achieves low computational complexity and is thus computationally tractable.
    \item We compare our algorithm with that  from \cite{RZF_SOPC} in terms of computational time and achieved user received power.
    \item We numerically show that, thanks to the low-dimensional representation of the LER, our solution presents a better mitigation of the received power over the continuous LER compared to existing methods.
\end{itemize}

\section{Signal model} 
\begin{figure}[!t]
    \centering
    \includegraphics[width=0.5\linewidth]{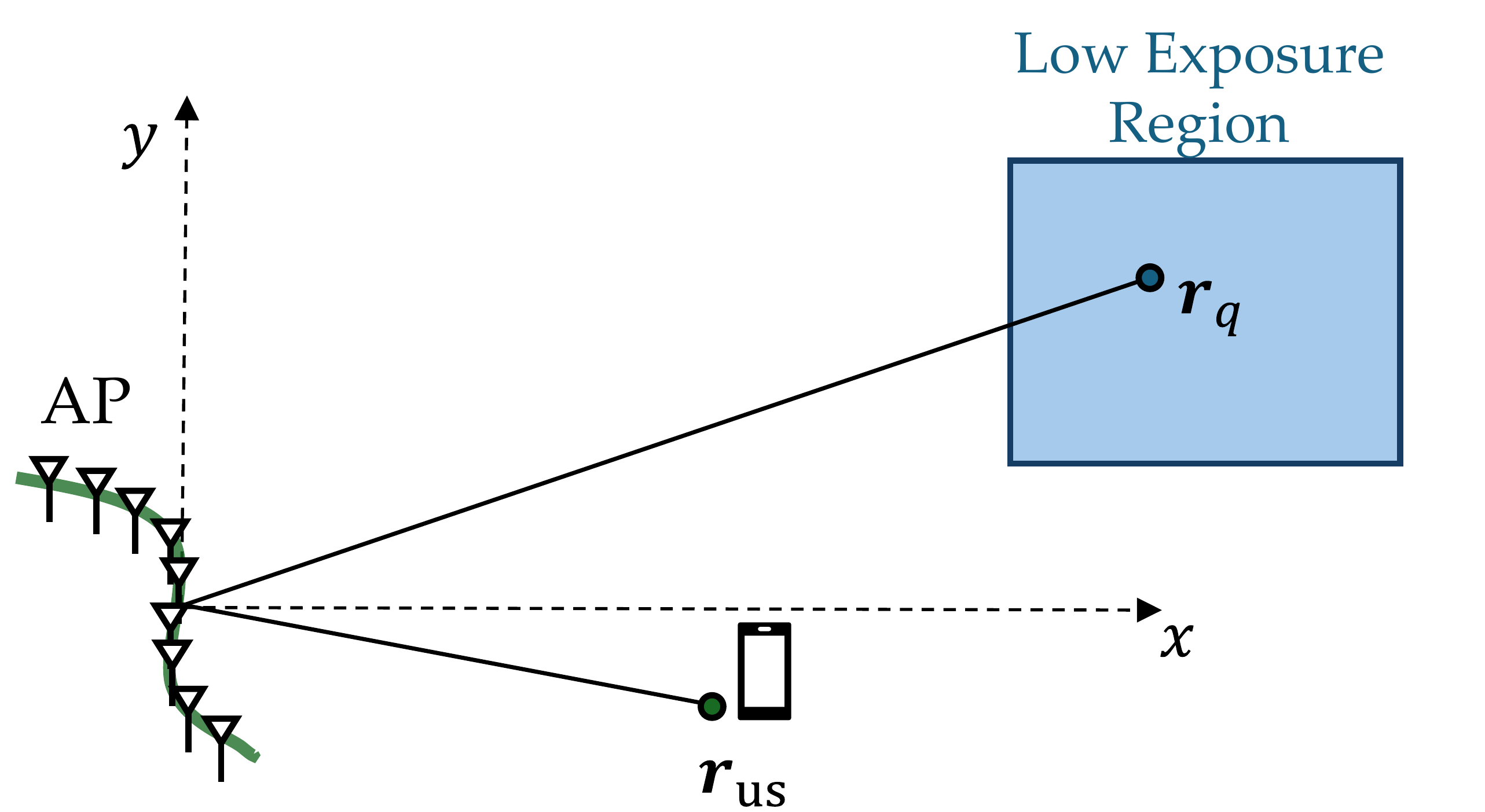}
    \caption{Scenario}
    \vspace{-5mm}
    \label{fig:Scenario}
\end{figure}

We consider a two-dimensional scenario in the $x-y$ plane, where an access point (AP) composed of an antenna array serves one user. The AP is centered at the origin, as shown in Fig. \ref{fig:Scenario}. It is composed of $N$ antennas separated by half a wavelength.  Let $\mathbf{r_\textrm{us}}$ be the location of the user. In this scenario, the user is assumed to lie in the NF region of the AP.

The steering vector corresponding to a position $\mathbf{r}$ is denoted as $\mathbf{a}(\mathbf{r})$, and the $n^{th}$ entry of the steering vector is given by 
\begin{equation}
    \left[\mathbf{a}(\mathbf{r} )\right]_n = \frac{1}{\sqrt{N}} e^{-j \frac{2\pi}{\lambda} \| \mathbf{r}-\mathbf{x}_n \|_2 },
\end{equation}
where $\lambda$ is the wavelength, $\mathbf{x}_n$ is the position of the $n^{th}$ antenna, and the normalization factor $1/\sqrt{N}$ ensures a received power independent of the number of antennas $N$. The steering vector of the user of interest is denoted as $\mathbf{a}_\textrm{us} \triangleq \mathbf{a}( \mathbf{r_\textrm{us}})$. The attenuation of the channel is omitted, and we only consider line-of-sight. Therefore, if a narrowband signal $s(t)$ with unit power is sent in downlink  with a precoding vector $\mathbf{w}^*$, the signal received at position $\mathbf{r}$ is given by
\begin{equation}
    y(t) = \mathbf{w}^H \mathbf{a}(\mathbf{r} ) s(t) + n(t),
\end{equation}
where  $n(t)$ is an additive white Gaussian noise. The useful power received at position $\mathbf{r}$ is given by
\begin{equation}
    P( \mathbf{r}) = \left| \mathbf{w}^H \mathbf{a}( \mathbf{r}) \right|^2.
\end{equation}
The power received by the user is denoted as $P_\textrm{us} \triangleq P( \mathbf{r_\textrm{us}})$.
To represent the characteristics of the LER, the latter is uniformly and densely sampled with a finite number $Q$ of positions, which typically verifies $Q>N$. The $q^{th}$ position is denoted as $\mathbf{r}_q$, and $\mathbf{a}_q \triangleq \mathbf{a}(\mathbf{r}_q)$ is its steering vector. We define the matrix $\mathbf{A} = [\,\mathbf{a}_1 \; \mathbf{a}_2 \; \cdots \; \mathbf{a}_Q\,] \in \mathbb{C}^{N\times Q}$.

\section{Precoder design}

The precoder design aims to maximize $P_\textrm{us}$ while ensuring that the power received at any position within the LER remains below a given threshold $t$. To address the challenging continuous constraint, this paper adopts an approach based on the discretization of the region through a sampling grid. The choice of the sampling step is discussed in Section \ref{sec:Num_Res}. The LER is therefore represented by the steering vectors associated with the $Q$ sampled positions, and the resulting optimization problem, denoted as P1, is a relaxed formulation of the continuous problem. This relaxed problem is given by 
\begin{equation}\label{Opti discret}\tag{P1}
\begin{aligned}
    \widehat{\mathbf{w}}
    &= \underset{\mathbf{w}  \in \mathbb{C}^N}{\operatorname{argmax}} \; \bigl| \mathbf{w}^H \mathbf{a}_\textrm{us} \bigr|^2 \\
    \text{s.t.} \quad 
    & \bigl| \mathbf{w}^H \mathbf{a}_q \bigr|^2 \le t, \quad \forall q =1, 2, ..., Q, \\
    & \| \mathbf{w} \|_2^2 \le 1,
\end{aligned}
\end{equation}
where $\widehat{\mathbf{w}}$ is an optimal precoding vector. The last constraint limits the power of the precoding vector.

\subsection{Optimal precoder}
Problem P1 is not convex, as it involves the maximization instead of minimization of a convex objective function under convex constraints. Nevertheless, the optimal solution of P1 is invariant to phase-shifts, and this phase invariance can be exploited to reformulate the problem into an equivalent convex problem, denoted as P2 \cite{InvariantPhase_SOCP}. The solutions of P1 can be computed by solving P2, which is given by 
\begin{equation}\label{Opti SOCP}\tag{P2}
\begin{aligned}
    \widehat{\mathbf{w}}
    &= \underset{\mathbf{w}  \in \mathbb{C}^N}{\operatorname{argmax}} \; \operatorname{Re} \{  \mathbf{w}^H \mathbf{a}_\textrm{us} \} \\
    \text{s.t.} \quad 
    & \bigl| \mathbf{w}^H \mathbf{a}_q \bigr| \le \sqrt{t}, \quad \forall q= 1, 2, ..., Q, \\
    & \operatorname{Re} \{  \mathbf{w}^H \mathbf{a}_\textrm{us} \} \geq 0 \\
    & \operatorname{Im} \{  \mathbf{w}^H \mathbf{a}_\textrm{us} \} = 0 \\
    & \| \mathbf{w} \|_2^2 \le 1.
\end{aligned}
\end{equation}
Problem P2 is expressed as a second-order cone program (SOCP), and can be solved using interior-point (IP) methods for example. However, the number of constraints in this scenario can typically reach several thousands since the sampled positions must be dense enough and must cover the entire LER. In such a scenario, the complexity of classical solvers makes the computation of the optimal precoding vector intractable.

Therefore, this paper proposes a heuristic based on linear algebra that approximates the optimal solution of P1 while reducing the complexity, enabling tractable computation. The optimal solution of the convex reformulation P2, computed with an IP solver, is used as a benchmark to assess the performance of the proposed heuristic.

To motivate our approach, we first analyze the problem when $t=0$. In that case, the constraints of the LER simplify to $\mathbf{A}^H\mathbf{w} = \mathbf{0}$, which corresponds to a ZF beamformer \cite{NF_MU}, given by the projection of $\mathbf{a}_\textrm{us}$ onto the nullspace of $\mathbf{A}$:
\begin{equation}
    \mathbf{a}_{\bot} = \left(\mathbf{I} - \mathbf{\Pi_A}  \right)\mathbf{a}_\textrm{us},
\end{equation}
\begin{figure}[h!]
    \centering
    \includegraphics[width=0.5\linewidth]{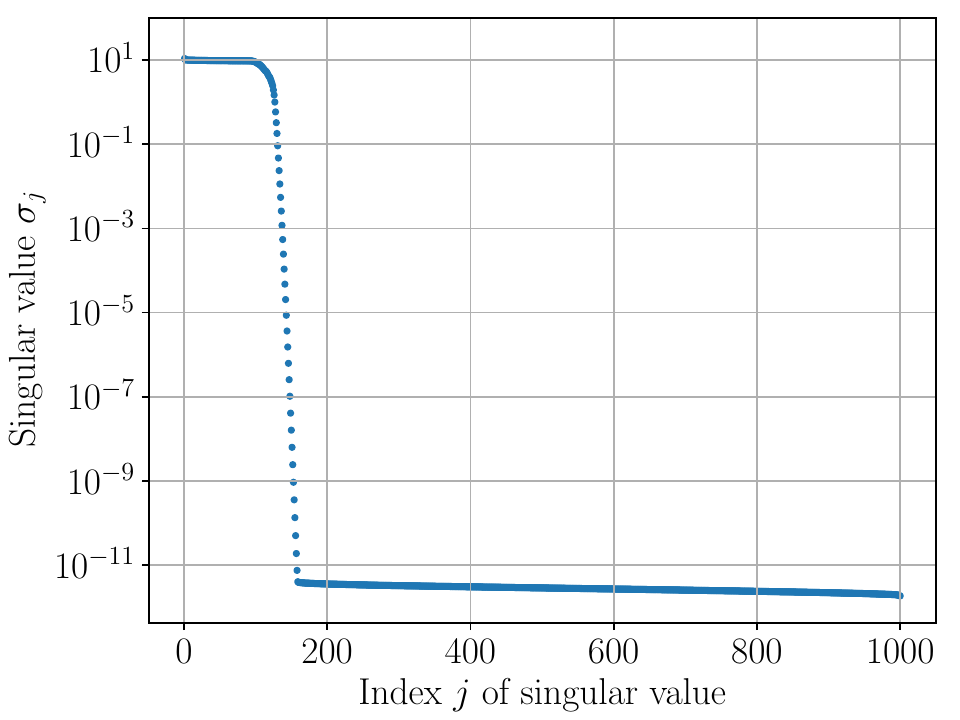}
    \caption{Singular values of $\mathbf{A}$ in a decreasing order, for a rank $r=N=1000$.}
    \label{fig:Singular_values}
\end{figure}
with $\mathbf{\Pi_A} = \mathbf{A} \left( \mathbf{A}^H \mathbf{A} \right)^{-1} \mathbf{A}^H $ being the projection matrix into the column space of $\mathbf{A}$, denoted as $\mathcal{C}_\mathbf{A}$. To meet the unit norm constraint, the ZF precoding vector is given by 
\begin{equation}
    \widehat{\mathbf{w}}_{\textrm{ZF}} = \frac{\mathbf{a}_{\bot}}{\|  \mathbf{a}_{\bot} \|}.
\end{equation}
In this scenario, as the number of constraints is higher than the number of antennas, matrix $\mathbf{A}$ is full-rank with a probability asymptotically equal to 1, leading to the rank $r$ equal to $r=N$. In that case, the space $\mathcal{C}_\mathbf{A}$ is the entire space $\mathbb{C}^N$ and the solution orthogonal to $\mathcal{C}_\mathbf{A}$ falls to the null vector. In such conditions, a non-null solution of P1 exists if and only if the threshold $t$ is non-zero, which corresponds to a RZF \cite{RZF_SOPC}. With a threshold different than zero, our heuristic exploits the structure of $\mathbf{A}$ to approximate the solution of P1 with low complexity. The proposed approach is based on a reduced-dimensional representation of the constraints and is detailed in the following section.
\subsection{Dominant subspace projection approach}
As the $Q$ positions are densely sampled from a compact CR in the NF region of the AP, the associated steering vectors are strongly correlated. This strong correlation induces a concentration of most of the energy of $\mathcal{C}_\mathbf{A}$
into a small number of singular components. These dominant components can be computed with a singular value decomposition (SVD), given by $\mathbf{A} = \mathbf{B} \mathbf{\Sigma} \mathbf{V}^H$. The most significant components of $\mathcal{C}_\mathbf{A}$ are identified by ordering the basis vectors $\mathbf{b}_j, \forall j=1, 2, ..., N$ according to decreasing singular values. For a rectangular LER of $500\lambda \times 500\lambda$, located $2000\lambda$ away from the AP and sampled every $5\lambda$ for each dimension, the corresponding singular values are illustrated in
Fig. \ref{fig:Singular_values}. It can be observed that due to the correlation of the steering vectors inside the compact region, more than $80\%$ of the singular values are negligible.

Based on this observation, we construct a low-dimensional subspace $\mathcal{C}_\mathcal{S}$ that approximates $\mathcal{C}_\mathbf{A}$ by retaining only a subset $\mathcal{S}$ of dominant singular components of $\mathcal{C}_\mathbf{A}$, identified via the SVD. This choice is motivated by the fact that the SVD provides the optimal low-rank approximation of $\mathbf{A}$, meaning that the projection of $\mathcal{C}_\mathbf{A}$ onto $\mathcal{C}_\mathcal{S}$ minimizes the least-squares approximation error. We name our heuristic presented in this section the \textit{Dominant Subspace Projection (DoSP)}.

Without the heuristic, thus using all the basis vectors of $\mathcal{C}_\mathbf{A}$, the conventional ZF solution can be rewritten as the normalization of the perpendicular vector
\begin{equation}\label{New_ZF}
    \mathbf{a}_{\bot} = \mathbf{a}_\textrm{us} - \sum_{j=1}^{r}\left(\mathbf{b}_j^H\mathbf{a}_\textrm{us}\right)\mathbf{b}_j.
\end{equation}
In \eqref{New_ZF}, the term $\sum_{j=1}^{r}\left(\mathbf{b}_j^H\mathbf{a}_\textrm{us}\right)\mathbf{b}_j$ represents the projection of $\mathbf{a}_\textrm{us}$ onto $\mathcal{C}_\mathbf{A}$, which is computed as the sum of the projections onto each basis vector of $\mathcal{C}_\mathbf{A}$. In our heuristic, $\mathcal{C}_\mathbf{A}$ is approximated by $\mathcal{C}_\mathcal{S}$, and the projection of $\mathbf{a}_\textrm{us}$ is computed by summing only the projections onto the basis vectors that belong to $\mathcal{S}$. The precoding vector is therefore given by
\begin{align}\label{w_s}
     \widehat{\mathbf{w}}_\mathcal{S} &= \frac{\mathbf{a}_\textrm{proj}}{\|  \mathbf{a}_\textrm{proj} \|},\\
    \mathbf{a}_\textrm{proj} &= \mathbf{a}_\textrm{us} - \sum_{j \in \mathcal{S}}\left(\mathbf{b}_j^H\mathbf{a}_\textrm{us}\right)\mathbf{b}_j.
\end{align}
Unlike $\mathcal{C}_\mathbf{A}$, the dimension of $\mathcal{C}_\mathcal{S}$ is lower than $N$ since it is composed of a subset $\mathcal{S}$ of the basis vectors, which enables a non-zero solution denoted as $\widehat{\mathbf{w}}_\mathcal{S}$. By construction, $\widehat{\mathbf{w}}_\mathcal{S}$ is orthogonal to $\mathcal{C}_\mathcal{S}$, but it is not strictly orthogonal to $\mathcal{C}_\mathbf{A}$. As a consequence, the power received in the LER is non-zero, which is acceptable as long as the maximum power received among the $Q$ positions in the LER remains below the threshold $t$. We define $k$ as the number of dominant basis vectors included in $\mathcal{S}$. The key challenge is to determine the optimal number $k_\textrm{opt}$ which maximizes $P_\textrm{us}$ while meeting the LER constraint. With the precoding vector given in \eqref{w_s}, the power $P_\textrm{us}$ and the power received at position $q$ in the LER can be expressed, respectively, by 
\begin{align}
    \left|\widehat{\mathbf{w}}_\mathcal{S}^H \mathbf{a}_\textrm{us} \right|^2 &= 1 - \sum_{j \in \mathcal{S}} \left| \mathbf{b}_j^H \mathbf{a}_\textrm{us}  \right|^2, \label{equa power set S user} \\
    \left|\widehat{\mathbf{w}}_\mathcal{S}^H \mathbf{a}_q \right|^2 &= \frac{\left|\sum_{j\in \{1, 2, ... r\} \backslash \mathcal{S}}\left(\mathbf{b}_j^H\mathbf{a}_\textrm{us}\right)\sigma_j\left[\mathbf{v}_j\right]_q\right|^2}{1 - \sum_{j\in \mathcal{S}} \left| \mathbf{b}_j^H\mathbf{a}_\textrm{us} \right|^2},\label{equa power set S LER}
\end{align}
where $\sigma_j$ is the singular value of the $j^{th}$ basis vector, and $\left[\mathbf{v}_j\right]_q$ is the $q^{th}$ entry of the $j^{th}$ row of $\mathbf{V}$. We observe from \eqref{equa power set S user} and \eqref{equa power set S LER} that as the number $k$ of vectors in $\mathcal{S}$ decreases, the power $P_\textrm{us}$ can only increase, and the power in the LER has a dominant tendency to increase.

\begin{figure}[!t]
    \centering
    \includegraphics[width=0.60\linewidth]{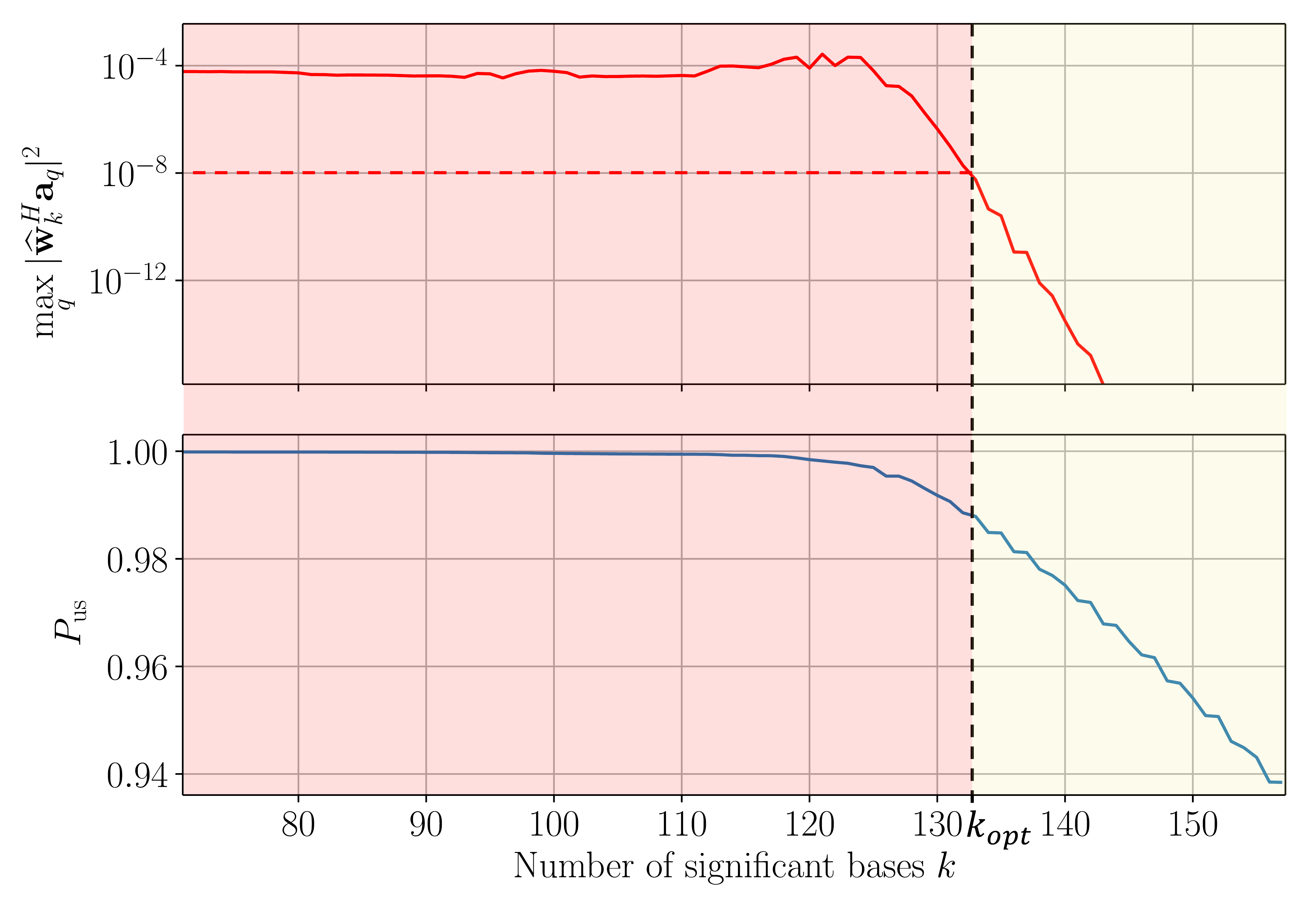}
    \caption{The power $\underset{q}{\max} \: |\widehat{\mathbf{w}}_k^H\mathbf{a}_q|^2$ is the maximum power (compared to MRT) received across all sampled positions of the LER, with a precoding vector $\widehat{\mathbf{w}}$ computed with $k$ basis vectors. In this example, $t=-80$ dB compared to MRT, and the start value is $k_\textrm{init}=156$.}
    \label{fig:k_both}
\end{figure}

The optimal number $k_\textrm{opt}$ of dominant basis vectors to include in $\mathcal{S}$ is determined through a decreasing search procedure. Starting from an initial value $k_\textrm{init}$, we iteratively decrease $k$ and evaluate if constraint \eqref{equa power set S LER} is met for all LER positions. As seen in Fig. \ref{fig:k_both}, the maximum power among the LER positions is monotonically decreasing with $k$ for a sufficiently high $k$. Since $k$ is initialized in that regime, we reduce $k$ until the LER constraint is violated. The last valid value yields $k_{\text{opt}}$, which maximizes $P_\textrm{us}$ while satisfying the LER constraint. To reduce computational complexity, $k$ is not initialized at $k_{\text{init}} = N$. Instead, we exploit the observation from Fig. \ref{fig:Singular_values} that a significant portion of the singular values are negligible. 
We discard the negligible singular values that are lower than a threshold $\sigma_{\text{th}} = 10^{-10} \sigma_1$, with $\sigma_1$ being the largest singular value. This yields an initial rank $k_{\text{init}} \ll N$, significantly reducing the number of iterations required to find $k_{\text{opt}}$.
The threshold $\sigma_{\text{th}}$ is chosen to be sufficiently small to ensure that the first iteration with $k_{\text{init}}$ meets the LER constraint, while remaining above the numerical precision. This condition is easily verified for the values of $t$ considered in this work.

As observed in Fig. \ref{fig:k_both}, if $k$ is lower than $k_\textrm{opt}$, then the space $\mathcal{C}_\mathcal{S}$ is too small and insufficiently captures the LER constraints. As a result, the maximum power received in the LER is higher than $t$ and at least one constraint is violated. If $k$ is higher than its optimal value, then the LER constraint is still met, but $P_\textrm{us}$ is not maximized.

\section{Numerical results}\label{sec:Num_Res}
Simulations are conducted for a scenario with an AP composed of a vertical uniform linear array of $N=1000$ antennas, corresponding to a dimension $D=500\lambda$ centered around the origin. The LER is a square with dimensions $500\lambda \times 500\lambda$, located $2000\lambda$ away from the AP. The user is located at Cartesian coordinates $(2200\lambda, -200\lambda)$, and the threshold $t$ is set to $-80$ dB compared to the MRT gain equal to 0dB. The MRT gain corresponds to the maximum power achievable at the user without the LER constraint. The dimensions are expressed in terms of wavelength and are therefore frequency-independent. Nevertheless, for a carrier frequency of 30 $GHz$, it corresponds to an LER of 25m$^2$ comparable to the dimensions of a room, a distance of 20m (resp. 22m) between the AP and the LER (resp. the user), and a 5m-long AP. Such an antenna array can be realized with radio stripes placed on a room's wall, such as in \cite{RadioStripes}.

The sampling step in the LER is $\Delta=5\lambda$ in both dimensions, yielding $Q=10201$ constraints, approximately ten times more than the number of antennas. The value of $\Delta$ is determined empirically to ensure that the power received on positions outside the sampling grid within the LER remains below a tolerance $\alpha = 10t$ when the precoding vector is the optimal solution of the relaxed discrete problem P1. We observe that the acceptable value of $\Delta$ depends on the resolution of the ambiguity function evaluated around the LER location, which depends on the AP size and the distance between the LER and the AP.

The performance metrics are: (i) the achievable $P_\textrm{us}$, (ii) the maximum power received in the continuous LER (i.e., for any position regardless of the sampling grid), and (iii) the computational complexity, assessed through the numerical computation time. The performance of the proposed DoSP method is compared to that achieved with an optimal IP solver, and to the performance of the algorithm proposed in \cite{RZF_SOPC}. 

The authors in \cite{RZF_SOPC} present an algorithm to compute RZF beamforming with a reduced complexity by solving the same optimization problem P1. Their algorithm relies on the assumption of linear independence, which does not hold in our scenario with highly correlated constraint steering vectors. Nevertheless, since \cite{RZF_SOPC} proposes a different suboptimal algorithm to solve the same problem P1, its performance is evaluated in our scenario and compared with that of DoSP to highlight the benefits of exploiting the high correlation among the LER steering vectors.

\subsection{Performance for the discrete problem P1}

\begin{figure}[!t]
    \centering
    \begin{subfigure}[t]{0.4\linewidth}
        \centering
        \includegraphics[width=1.0\linewidth]{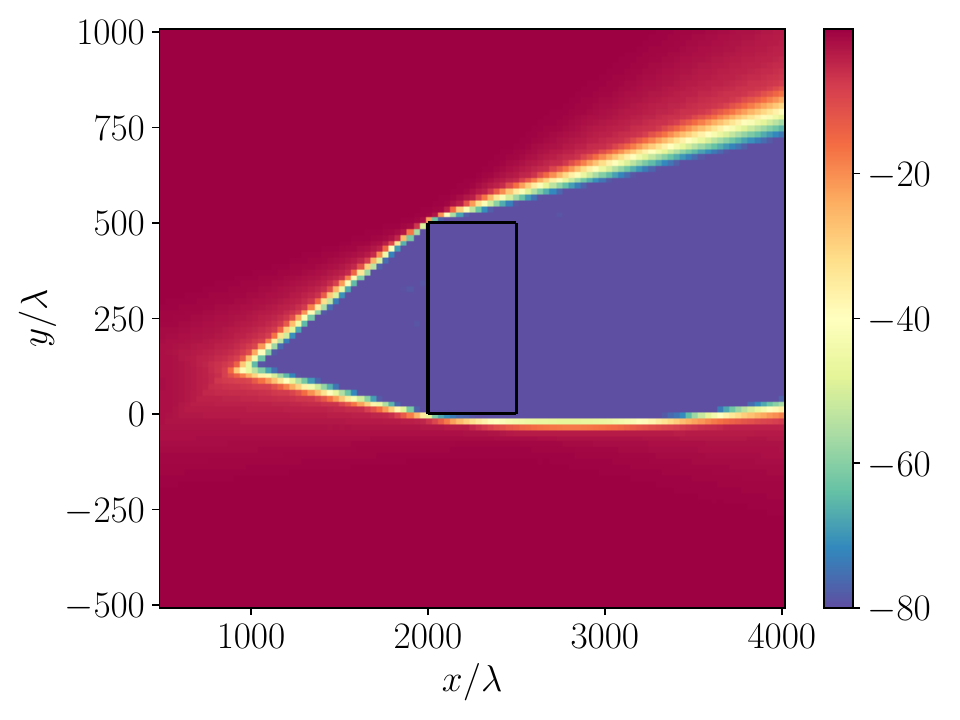}
        \caption{LER of 25m$^2$.}
        \label{power_zone1}
    \end{subfigure}%
    ~ 
    \begin{subfigure}[t]{0.4\linewidth}
        \centering
        \includegraphics[width=1.0\linewidth]{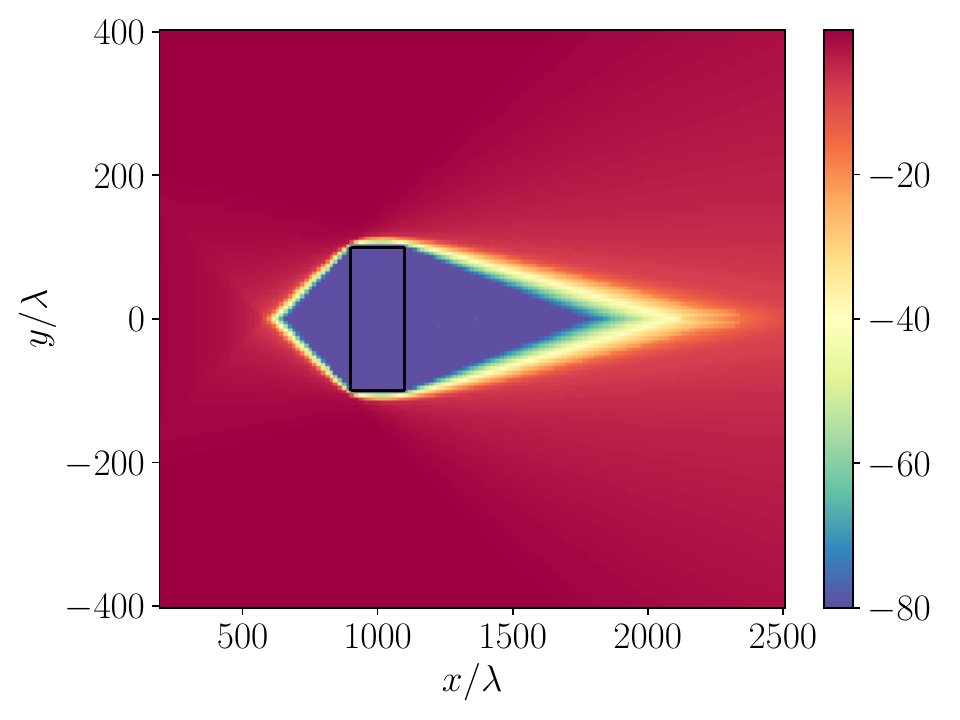}
        \caption{LER of 4m$^2$.}
        \label{power_zone2}
    \end{subfigure}
    \caption{Achievable $P_\textrm{us}$ compared to MRT (in dB), with respect to the user position. The LER is indicated in black and is located in $[2000\lambda; 2200\lambda] \times [0; 500\lambda]$ (left) and $[900\lambda; 1100\lambda] \times [-100\lambda; 100\lambda]$ (right).  }
    \label{Achievable_Power}
\end{figure}

First, DoSP is compared in Table \ref{tab:compBis} with the IP solver which provides the exact optimizer of P2 and with \cite{RZF_SOPC}, based on the achieved $P_\textrm{us}$ and computation time.
\begin{table}[!t]
\caption{Comparison between DoSP, optimal solver and \cite{RZF_SOPC}.}
\begin{center}
\begin{tabular}{|c|c|c|c|}
\hline
 &Optimal& DoSP& \cite{RZF_SOPC}\\
\hline
$P_\textrm{us}$ compared to MRT &  99.2\% & 98.5\% &  97.5\% \\
\hline
Precompute time & 0s & 25s & 0s \\
\hline
Real-time compute time & 1h & 0.7s & 27s\\
\hline
\end{tabular}
\label{tab:compBis}
\end{center}
\end{table}
\begin{figure}[!t]
    \centering
    \begin{subfigure}[t]{0.4\linewidth}
        \centering
        \includegraphics[width=1.0\linewidth]{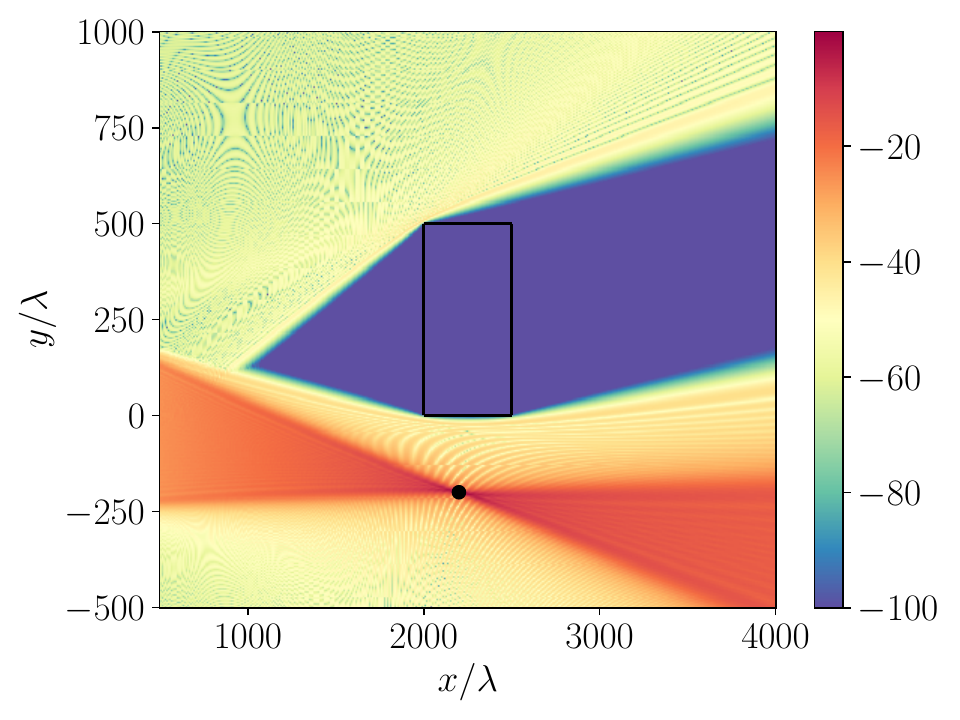}
        \caption{DoSP}
        \label{interf_Heuristic}
    \end{subfigure}%
    ~ 
    \begin{subfigure}[t]{0.4\linewidth}
        \centering
        \includegraphics[width=1.0\linewidth]{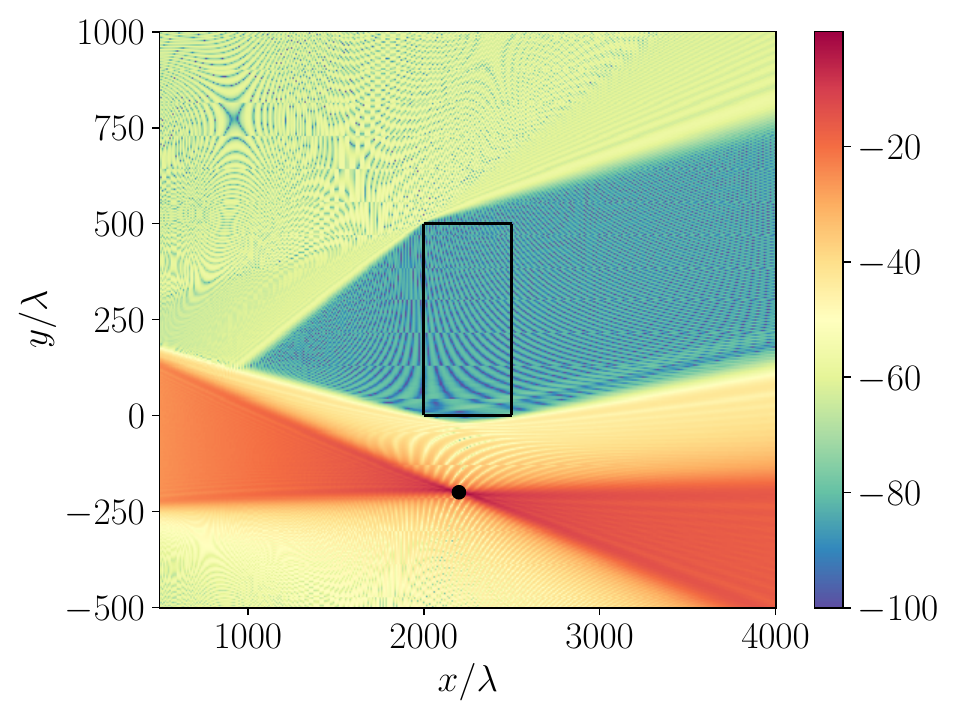}
        \caption{Optimal solver}
        \label{interf_Solver}
    \end{subfigure}
    \caption{Beam pattern compared to MRT (in dB). The user location and the LER delimitation are indicated as the black dot and black rectangle, respectively.}
    \label{Interf_Figures}
\end{figure}
It can be seen in Table \ref{tab:compBis} that $P_\textrm{us}$ achieved with DoSP is suboptimal but very close compared to optimality. Moreover, the computational complexity is negligible compared to numerical solvers, and is also more than one order of magnitude smaller than the solution proposed in \cite{RZF_SOPC}. 
Furthermore, the power achieved by \cite{RZF_SOPC} is slightly lower than with DoSP, as their method is designed for linearly independent interference channels rather than the highly correlated steering vectors associated with the compact LER.
The corresponding beam patterns achieved with the optimal solver and with DoSP are shown in Fig. \ref{Interf_Figures}. We observe that the received power is maximized at the user and remains lower than or equal to $t$ inside the LER, but also in a region outside the LER. The characterization of the exact geometry of that region is left as future work, but a mathematical explanation can be provided: the points in that region are associated to steering vectors that are approximately contained in $\mathcal{C}_\mathcal{S}$ (i.e., the vectors are approximately equal to their projection onto $\mathcal{C}_\mathcal{S}$). If the user position is such that $ \mathbf{a}_\textrm{us} \in \mathcal{C}_\mathcal{S}$, therefore the maximum possible $P_\textrm{us}$ is bounded by the threshold $t$. Fig. \ref{power_zone1} illustrates the maximum achievable $P_\textrm{us}$ with respect to the user position using DoSP, for the same LER as in Fig. \ref{Interf_Figures}. As expected, the achievable power is equal to $t$ if the user is located within the LER, but also if he/she is located in the positions outside the LER such that their steering vector still lies in $\mathcal{C}_\mathcal{S}$. The region of these positions presents the same geometry as in Fig. \ref{Interf_Figures}. That observation is inherent to the problem parameters and is independent of the precoding method. Fig. \ref{power_zone2} shows that, for a smaller LER, the NF properties allow a reasonable power coverage even for a user located between the AP and the LER, or behind the LER with respect to the AP.

\subsection{Received power in the continuous LER}
\begin{figure}[!t]
    \centering
    \includegraphics[width=0.5\linewidth]{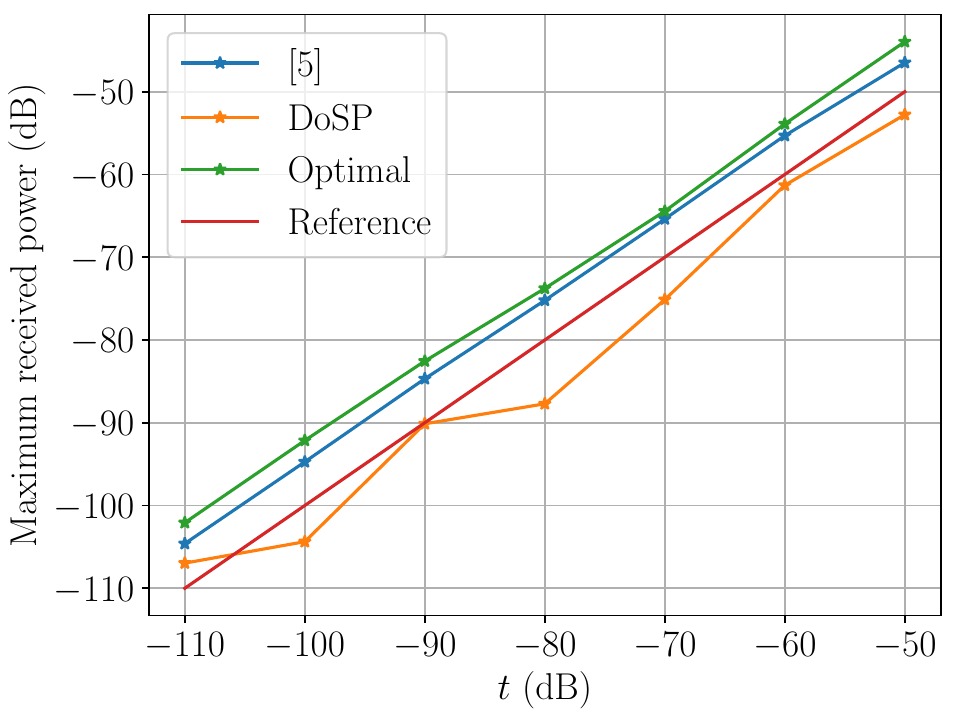}
    \caption{Maximum received power observed in the LER on a grid with spacing $\Delta_c=\lambda/4$. The red curve represents the reference threshold that the received power should not exceed.}
    \label{fig:MaxStopband_vsT}
\end{figure}

The optimal numerical solver, as well as the precoding vector from \cite{RZF_SOPC} and the DoSP beamformer, all compute an optimal solution of the relaxed problem P1 whose constraints are only met at the finite number of points in the LER. For the continuous LER, the received power at positions that do not lie on the sampling grid might exceed the threshold $t$. Fig. \ref{fig:MaxStopband_vsT} shows for different values of $t$ the maximum power observed in the LER on a thinner grid with a smaller sampling step $\Delta_c=\lambda/4$.
By design of the sampling step, the maximum value of received power in the CR exceeds the threshold $t$ while remaining below the tolerance $\alpha=10t$ for the optimal solver, as expected. Interestingly, the same observation applies to the precoding vector from \cite{RZF_SOPC}. By contrast, the power achieved with DoSP remains mostly below the threshold.

This behavior can be explained by the choice of basis vectors in the DoSP method, which represent best the structure of the LER since the SVD provides the optimal low-rank approximation of $\mathcal{C}_\mathbf{A}$. Conversely, the optimal solution returned by the IP solver and by \cite{RZF_SOPC} account only for the sampled points and do not guarantee any power constraint at positions lying outside the sampling grid. As a consequence, the power received in the LER is more heterogeneous than with DoSP, as it can be seen in Fig. \ref{interf_Solver}. Increasing the sampling density may yield a smoother power distribution within the LER, but at the expense of an increased computational complexity. By contrast, DoSP requires fewer constraints, which decreases the computational complexity, while maintaining an homogeneous power in the continuous LER.

\section{Conclusions}
In this paper, by leveraging the spherical wavefront exhibited in the NF region, we design a beam pattern that maximizes the power received at a user while creating a two-dimensional LER. Our algorithm approximates the optimal value of the given optimization problem with a low complexity that enables tractable computation. Moreover, our approach using an SVD leads to a more controlled power in the continuous LER even with a finite number of sampled positions.

\printbibliography

\end{document}